\documentclass{appolb}
\usepackage{epsfig}
\usepackage{graphicx}

\newcommand{\srt}{$\sqrt{s_{_{\rm NN}}}$}

\newcommand{\pt}{$p_T$ }

\begin{document}
\def\Journal#1#2#3#4{{#1} {\bf #2} (#4) #3}

\def\NCA{\em Nuovo Cimento}
\def\NIM{\em Nucl. Instr. Meth.}
\def\NIMA{{\em Nucl. Instr. Meth.} A}
\def\NPB{{\em Nucl. Phys.} B}
\def\NPA{{\em Nucl. Phys.} A}
\def\PLB{{\em Phys. Lett.}  B}
\def\PRL{{\em Phys. Rev. Lett.}}
\def\PRC{{\em Phys. Rev.} C}
\def\PRD{{\em Phys. Rev.} D}
\def\ZPC{{\em Z. Phys.} C}
\def\JPG{{\em J. Phys.} G}
\def\EPJ{{\em Eur. Phys. J.} C}
\def\RPP{{\em Rep. Prog. Phys.}}

\title{Open charm production at RHIC}

\author{Zhangbu Xu
\address{Physics Department, Brookhaven National Laboratory, Upton, NY 11973, USA}
}
\maketitle
\begin{abstract}

In this report, we present the measurements of open charm production
at mid-rapidity in p+p, d+Au, and Au+Au collisions at RHIC
energies. The techniques of direct reconstruction of open charm via its hadronic 
decay and indirect measurements via its semileptonic decay are
discussed. The beam energy dependence of total charm cross section,
electron \pt spectra, and their comparisons to theoretical
calculations, including NLO pQCD, are presented. The electron spectra
in p+p, d+Au, and Au+Au collisions at \srt=200 GeV show significant
variation. The open charm absolute cross section at midrapidity and
its centrality dependence are compared to those of inclusive hadrons
integrated over $p_{T}>1.5$ GeV/c.

\end{abstract}

\section{Introductions}

Hadrons with heavy flavor are unique tools for studying the strong
interaction described by Quantum Chromodynamics (QCD). Due to the
large charm quark mass, which requires large energies ($^{>}_{\sim}3$
GeV) for their creation, charm quark production can be evaluated by
perturbative QCD (pQCD) even at low momentum with the introduction of
additional scales related to its mass~\cite{charm1,charm2}. Therefore,
the theoretical calculations of the charm hadron production cross
section integrated over momentum space are expected to be less
affected by non-perturbative processes and hadronization than
those of the light hadrons~\cite{phenixpi0}. Charm
production has been proposed as a sensitive measurement of
parton(gluon) distribution function in nucleon and the nuclear
shadowing effect by systematical studies of p+p, and p+A 
collisions~\cite{lin96}. The reduced energy loss of heavy
quarks(``deadcone'' effect) at momentum range
$5^{<}_{\sim}p_{T}{}^{<}_{\sim}10$ GeV/c will help us study the
energy loss mechanism within the partonic
medium~\cite{dokshitzer01}. A possible enhancement of charmonium
($J/\psi$) production can be present at RHIC energies
\cite{pbm,rafelski,mclerran} through charm quark coalescence. This
effect is opposed to the $J/\psi$ suppression in a Quark-Gluon
Plasma(QGP) in the absence of that process~\cite{matsui}. The
measurement of both open charm yield and charmonia may allow us to
quantify the effects and study whether charms are in chemical
equilibrium with the system. In addition, the heavy flavor transverse
momentum distributions and their anisotropic flow can be used to study
the nature of early thermalization in A+A collisions~\cite{xu2}. 

Identification of charmed hadrons is difficult due to its short
lifetime ($c\tau(D^{0})=124$ $\mu$m), low production rate and
overwhelming combinatoric background. Most measurements of the total
charm cross section in hadron-hadron collisions were performed at low
center-of-mass energies ( ${}^{<}_{\sim}$ 40 GeV) in fixed target
experiments~\cite{tavernier,fermic}. The measurements at high energy
colliders were either at high $p_{T}$~\cite{CDFII} , with large
uncertainty~\cite{phenixe,ua2} or
inconsistency~\cite{isrc}. Theoretical predictions for RHIC energy
region differ significantly~\cite{vogt02, dipole}. At RHIC, direct
measurements of open charm from charm hadronic decays and indirect
measurements from charm semileptonic decays in many beam conditions
are possible. We will illustrate the techniques used in open charm
measurements at RHIC and discuss the results.

\section{Direct open charm reconstruction}
\begin{figure}[h]
\centerline{\includegraphics[width=1.0\textwidth] {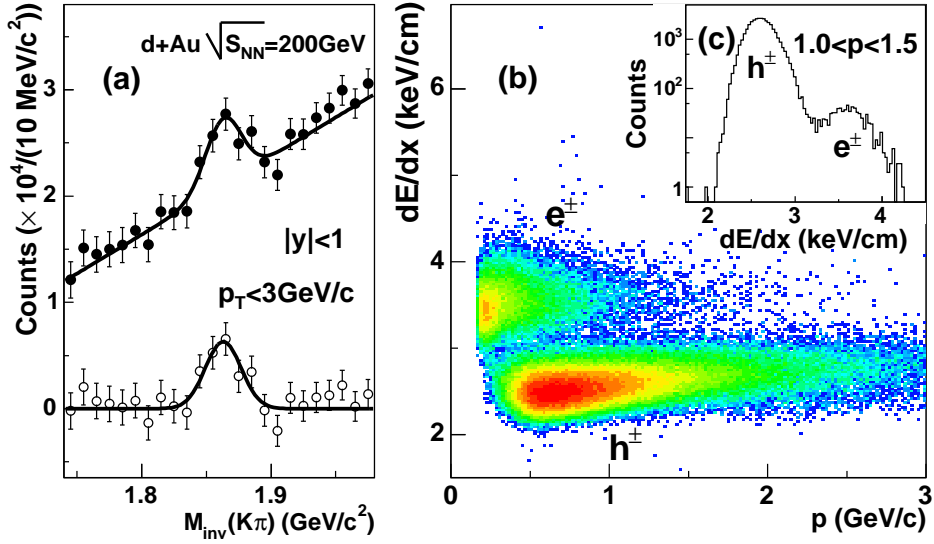}}
 \caption[]{(a) Invariant mass distributions of kaon-pion pairs from
 d+Au collisions. The solid circles depict signal after the
 mixed-event background subtraction, the open circles after
 subtraction of the residual background by a linear parameterization.
 (b) Ionization energy loss ($dE/dx$) in the TPC vs. particle momentum
 ($p$) with a TOF cut of $|1/\beta - 1| \le 0.03$.  Insert: projection
 on the $dE/dx$ axis for particle momenta 1$<p<$1.5 GeV/$c$. Figure
 from ~\cite{starcharm}.}
 \label{fig1pid}
\end{figure}

\begin{itemize}
\item{\bf Large acceptance, Large dataset}\\
The data used in $D^0$ direct reconstruction were taken during the
2003 RHIC run in d+Au collisions at \srt = 200 GeV with the Solenoidal
Tracker at RHIC (STAR)~\cite{starcharm,ruanlj}. A total of 15.7
million minimum bias triggered d+Au collision events were used in
$D^{0}$ analysis.  The primary tracking device of the STAR detector is
the Time Projection Chamber (TPC) which was used to reconstruct the
decay of $D^0\rightarrow K^-\pi^+$ ($\overline{D^0}\rightarrow
K^+\pi^-$) with a branching ratio of 3.83\%.
\item{\bf Event Mixing}\\ The invariant mass
spectrum of $D^0(\overline{D^0})$ was obtained by pairing oppositely
charged kaon and pion in same event with the parent rapidity 
$|y|<1$. The kaon and pion tracks were identified through the
ionization energy loss ($dE/dx$) in the TPC. The track reconstruction in
the TPC gives a single track projection resolution of $\sim 1$ cm
around the collision vertex and therefore does not allow to resolve
the exact decay topology.  The $D^0$ signal in $p_T<3$ GeV/$c$ and
$|y|<1$ after mixed-event background subtraction~\cite{kstarprc} is
shown in the left panel of Fig.~\ref{fig1pid}.
\end{itemize}
In current STAR analyses, the total charm cross section is largely
determined by directly reconstructed $D(\bar{D})$ at low
\pt~\cite{starcharm,ruanlj}.  In addition, $D^{*}$ has been measured at
$1.5^{<}_{\sim}p_{T}<6$ GeV/c and $D^{\pm}, D^{0}$ at $p_{T}\simeq 10$
GeV/c~\cite{antai}. In this presentation, we will focus on total charm 
cross section and its centrality dependence~\cite{starcharm,phenix200e}. 

\section{Electron spectra from charm semileptonic decay}
There are three techniques used to identify electrons in STAR and PHENIX: 
\begin{enumerate}
\item{\bf Tracking+RICH+EMCAL} \\ This method is used by PHENIX
collaboration to identify electrons~\cite{phenixe,phenix200e}.
Tracking from drift chamber and pad chamber provides particle track
with high momentum resolution. A Ring Image Cerenkov Detector (RICH)
is used as veto counter, and an electromagnetic calorimeter confirms
the track's existence and provides a E/p measurement for electron
selection.
\item{\bf Tracking+dE/dx+EMC}\\ This method is used by STAR
collaboration to identify electron with $p^{>}_{\sim}1.5$
GeV/$c$~\cite{starcharm,ruanlj}.  The STAR TPC not only provides
tracking in a solenoidal field, but also has good particle
identification capabilities through ionalization energy loss at
resolution of about 8\%. Due to relativistic rise of dE/dx of electron
at high $\beta\gamma$, its separation from $\pi,K$ and p are very good
at hadron rejection of $>10^{-2}$ at high \pt. The addition of E/p=1.0
from EMC further reject $\pi,K$ and p, and hadrons with high dE/dx
(nuclei, ghost tracks). Since EMC can be used as a trigger detector,
the electron spectra from this method can be extended to much higher
\pt with high statistics~\cite{alexi}.
\item{\bf Tracking+dE/dx+TOF}\\ This new technique is developed by
STAR collaboration.  A prototype time-of-flight system
(TOFr)~\cite{startof1} based on the multi-gap resistive plate chamber
technology was installed in STAR. It covers $-1<\eta<0$ and allows
particle identification for $p_{T}<3.5$ GeV/$c$. In addition to its
capability of hadron identification~\cite{startof1}, electrons could
be identified at low momentum ($0.2^{<}_{\sim}p_{T}<3$ GeV/$c$) by the
combination of velocity ($\beta$) from TOFr and $dE/dx$ from TPC
measurements. The right panel of Fig.~\ref{fig1pid} demonstrates the
clean separation of electrons from hadrons using their energy loss
($dE/dx$) in the TPC after applying a TOFr cut of $|1/\beta - 1| \le
0.03$. This cut eliminated the hadrons crossing the electron $dE/dx$
band.  Electrons were required to originate from the primary vertex.
Hadron contamination was evaluated to be about $10-15$\% in this
selection.
\end{enumerate}
\begin{figure}[h]
\centerline{\includegraphics[width=1.0\textwidth]{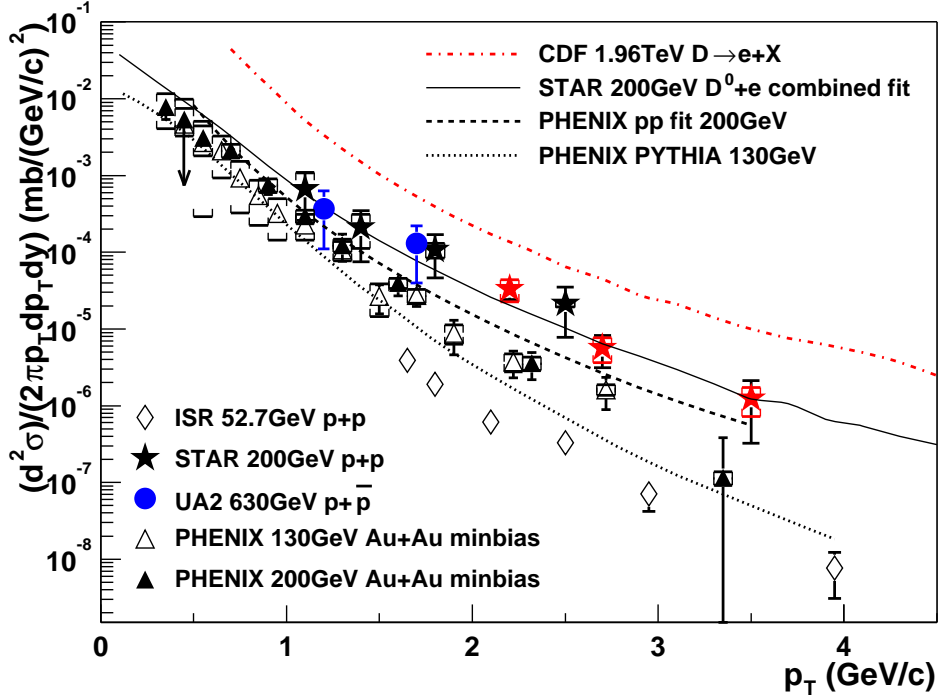}}
\caption[]{Non-photonic electron \pt distributions from p+p,
p+$\bar{p}$, d+Au,and Au+Au at different beam energies from ISR to
Tevetron.}
\label{fig3de}
\end{figure}
Gamma conversions $\gamma\rightarrow e^{+}e^{-}$ and
$\pi^0\rightarrow\gamma e^{+}e^{-}$ Dalitz decays are the dominant
photonic sources of electron background. There are again several
methods used to study the background electron sources:
\begin{enumerate}

\item{\bf Cocktail Modeling}\\ This method was adapted by PHENIX
Collaboration in their first publication~\cite{phenixe}.  The cocktail
is a detailed detector simulation of electrons/positrons from $\gamma$
conversion and Dalitz decays from $\pi^{0},\eta$ and other photonic
sources.
\item{\bf Converter}\\ This method was adapted by PHENIX Collaboration
in their latest report~\cite{phenix200e}. Two datasets with and
without a converter with known thickness were taken and analyzed. The
two measurements allows to untangle the sources associated with photon
conversion and those without. Species-dependent Dalitz fraction per
$\gamma$ and contribution from other photonic sources ($\eta, \omega,
\rho, \phi$ and $K$) are evaluated by detailed detector simulations.
\item{\bf Invariant Mass Reconstruction}\\ The $e^{+}e^{-}$ pairs from
photon conversions and Dalitz decays are present mainly at small pair
invariant mass and/or small opening angle~\cite{starcharm,ruanlj}. Due
to the large coverage of the TPC, the efficiency of finding pairs is
very high for such processes. To measure the background, the invariant
mass and opening angle of the pairs were constructed by first
selecting an electron (positron) in TOFr and then matching it with
every positron (electron) candidate reconstructed in the
TPC~\cite{johnson} without additional requirement of a secondary
vertex at the conversion point. In this method, both inclusive
electron spectra and background sources are taken from same
dataset, and the reconstructed background spectrum is not sensitive to
the detailed knowledge of the conversion and Dalitz decay. About 95\% of
electrons from sources other than charm semileptonic decays have been
measured with this method, while the remaining fraction ($<5\%$) from
decays of $\eta, \omega, \rho, \phi$ and $K$ was determined from
detailed detector simulations.
\end{enumerate}

\begin{figure}[h]
\centerline{\includegraphics[width=1.0\textwidth]{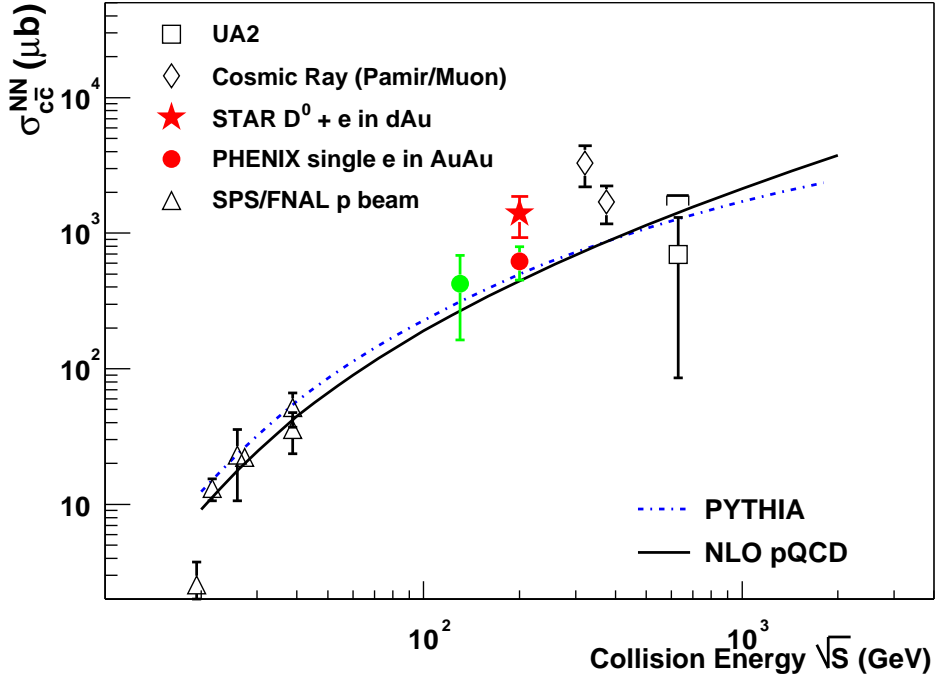}}
\caption[]{Total $c\bar{c}$ cross section per nucleon-nucleon
collision vs. the collision energy (in \srt). 
}
\label{fig4e}
\end{figure}

\section{Discussions}
\begin{figure}[h]
\centerline{\includegraphics[width=1.0\textwidth]{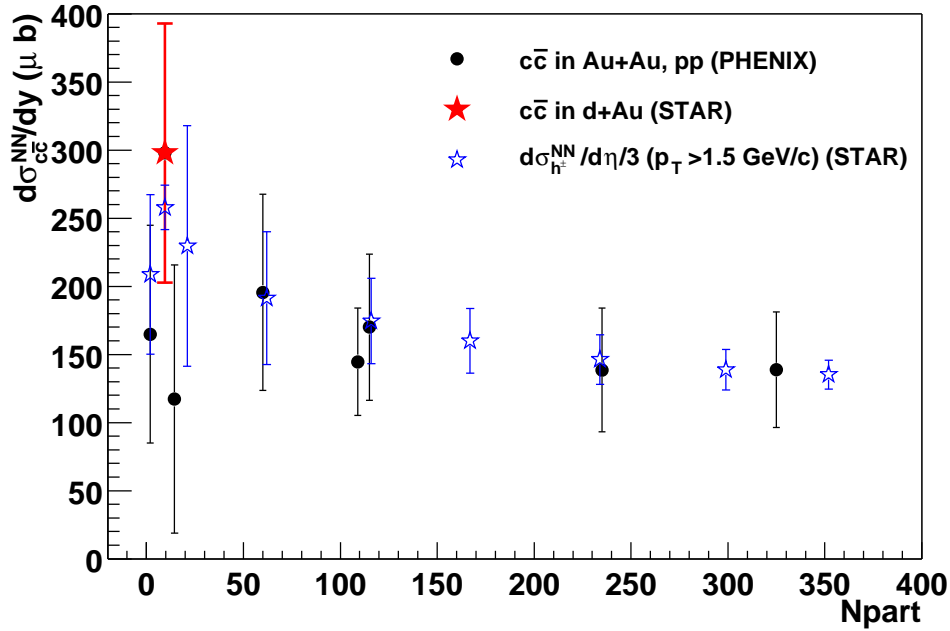}}
\caption[]{Charm differential cross section per binary nucleon-nucleon collision 
($d\sigma^{NN}_{c\bar{c}}/dy$)
at midrapidity as function of number of participant nucleons at RHIC
\srt=200 GeV. Also plotted as open stars are cross section per binary 
nucleon-nucleon collisions 
($d\sigma^{NN}_{h^{\pm}}/d\eta$ ($\mu b$)) of inclusive hadrons integrated
over $p_{T}>1.5$ GeV/c ($\simeq m_{D}$) and scaled down by a factor of three.
}
\label{fig5npart}
\end{figure}

Fig.~\ref{fig3de} shows the energy dependence of the electron \pt
spectra from ISR to Tevetron. Direct open charm measurements (STAR,
CDF) are converted to electron spectra. Several measurements at RHIC
have shown to be consistent between (dE/dx+EMC), direct {\bf D},
(dE/dx+TOF)~\cite{ruanlj,antai,alexi} at STAR, and to some extent, the
measurements in p+p collisions between STAR and PHENIX~\cite{ralfa}.
Fig.~\ref{fig3de} shows significant dispersion of electron
distribution at \srt=200 GeV at RHIC from Au+Au, d+Au to p+p at
$p_{T}\simeq2$ GeV/c where statistics are still high.  The variation
in spectra is possibly due to the interactions between the charm quark
(hadron) and the medium~\cite{teaney}. The deadcone
effect~\cite{dokshitzer01} would be most effective at
$5^{<}_{\sim}p_T{}^{<}_{\sim}10$ GeV/c and not yet accessible at the
\pt range ($\simeq2$ GeV/c) of electrons. The beam energy dependence
of the cross section is shown in Fig.~\ref{fig4e}. The PHENIX results
are derived from electron spectrum in Au+Au minbias collisions using
the electron spectrum shape from PYTHIA6.205~\cite{phenix200e}.  The
STAR result is from a combined fit of $D^{0}$ and electron spectra in
d+Au minbias collisions. At $\sqrt{s}\sim52-63$ GeV, the available
measurements are inconclusive due to the inconsistency between
different measurements~\cite{tavernier} and are omitted. At
$\sqrt{s_{_{NN}}}=200$ GeV, both PYTHIA and NLO pQCD calculations
underpredict the total charm cross section~\cite{vogt02,pythia1}.
This is evident from STAR data but less pronounced in PHENIX
results. There are indications that a large charm production cross
section ($\sigma_{c\bar{c}}^{NN}\simeq2-3$ mb) at
$\sqrt{s_{_{NN}}}\simeq300$ GeV is essential to explain cosmic ray
data~\cite{cosmic}.  The centrality dependence of charm $d\sigma/dy$
per binary nucleon-nucleon collision at midrapidity shows possible
dependence on centrality with lower production cross section in
central collisions as in Fig.~\ref{fig5npart}. In the same figure, the
cross sections per binary nucleon-nucleon collisions of inclusive
hadrons~\cite{starnch} ($d\sigma_{h^{\pm}}/d\eta/3$($\mu b$)), which
were integrated over $p_{T}>1.5$ GeV/c (close to the mass of charm
hadron $m_{D}$) and scaled down by a factor of 3 for three light quark
flavors, show same magnitude of absolute cross section and a similar
trend. This may suggest that particle production rate is not sensitive
to flavor when the momentum transfer is above the production
threshold.

A direct reconstruction of open charm spectra in Au+Au collisions from
$0^{<}_{\sim}p_{T}{}^{<}_{\sim}10$ GeV/c in future runs, when compared
with those from d+Au data~\cite{starcharm,antai}, will enable us to
study the thermalization at low \pt and deadcone effect at high \pt.
The elliptic flow $v_2$ measurements of charm flow with higher
statistics at the same \pt range will further strengthen the case,
since early thermalization will result in large $v_2$ at low \pt, and
absence of large energy loss due to deadcone effect may result in
smaller $v_2$ at high \pt when compared to those of light hadrons.
\section{Summary}
In summary, charm cross section and transverse momentum distributions
from p+p, d+Au and Au+Au collisions have been measured by the STAR and
PHENIX collaborations at RHIC. The NLO calculation underpredicts the
total cross section in this energy range. A possible centrality
dependence of charm cross section can explain the marginally different
cross sections observed by STAR in d+Au collisions and PHENIX in Au+Au
collisions. We look forward to more results using charm as probe to
the hot and dense medium created at RHIC from run4 and future runs.


\par {\bf Acknowledgement}: The author would like to thank X. Dong,
H. Huan, D. Kharzeev, A. Tai, T. Ullrich, N. Xu and H. Zhang for
valuable discussions. 



\begin{thebibliography}{999}
\bibitem{charm1} P.L. McGaughey {\it et al.},
{\em Int. J. Mod. Phys.} {\bf A 10}, 2999(1995).
\bibitem{charm2}M.L. Mangano {\it et al.}, \Journal{\NPB}{405}{507}{1993}.
\bibitem{phenixpi0} S.S. Adler {\it et al.}, (PHENIX
Collaboration), \Journal{\PRL}{91}{241803}{2003}; J. Adams {\it et
al.}, (STAR Collaboration), \Journal{\PRL}{92}{171801}{2004}.
\bibitem{lin96} Z. Lin and M. Gyulassy,
        \Journal{\PRL}{77}{1222}{1996}.
\bibitem{dokshitzer01} Y.L. Dokshitzer and D.E. Kharzeev,
        \Journal{\PLB}{519}{199}{2001}.
\bibitem{pbm} A. Andronic {\it et al.},
        \Journal{\PLB}{571}{36}{2003}, and references therein.
\bibitem{rafelski} R.L. Thews, M. Schroedter, and J. Rafelski,
        \Journal{\PRC}{63}{054905}{2001}, and references therein.
\bibitem{mclerran}  M.I. Gorenstein \etal, \Journal{\JPG}{28}{2151}{2002}.
\bibitem{matsui} T. Matsui and H. Satz,
        \Journal{\PLB}{178}{416}{1986}.
\bibitem{xu2} N. Xu and Z. Xu \Journal{\NPA}{715}{587c}{2003};
        S. Batsouli {\it et al.}, \Journal{\PLB}{557}{26}{2003};
    Z.W. Lin and D. Molnar, \Journal{\PRC}{68}{044901}{2003};
    X. Dong \etal, \Journal{\PLB}{597}{328}{2004}.
\bibitem{tavernier} S.P.K. Tavernier, \Journal{\RPP}{50}{1439}{1987};
and references therein.
\bibitem{fermic} G.A. Alves {\it et al.}, (E769 Collaboration),
        \Journal{\PRL}{77}{2388}{1996}.
\bibitem{CDFII} D. Acosta {\it et al.}, (CDF II Collaboration),
        \Journal{\PRL}{94}{241804}{2003}.
\bibitem{phenixe} K. Adcox {\it et al.}, (PHENIX Collaboration),
        \Journal{\PRL}{88}{192303}{2002}.
\bibitem{ua2} O. Botner {it et al.}, (UA2 Collaboration), \Journal{\PLB}{236}{488}{1990}.
\bibitem{isrc} F.W. B\"{u}sser {\it et al.}, \Journal{\NPB}{113}{189}{1976}.
\bibitem{vogt02} R. Vogt, hep-ph/0203151, and references therein.
The curve in Fig.~\ref{fig4e} is a NLO pQCD calculation with
CTEQ5M, $\mu_{F}=\mu_{R}=2m_{c}$, $m_{c}$ = 1.2 GeV/$c^{2}$.
\bibitem{dipole} J. Raufeisen and J.-C. Peng,
\Journal{\PRD}{67}{054008}{2003}.
\bibitem{starcharm}J. Adams, \etal (STAR Collaboration), nucl-ex/0407006. 
\bibitem{ruanlj} L.J. Ruan, \Journal{\JPG}{30}{S1197-S1200}{2004};
nucl-ex/0403054; H. Zhang, DNP 2003, Hot Quarks 2004.
\bibitem{kstarprc} C. Adler {\it et al.},
        {\em Phys. Rev.} {\bf C66}, 06190(R)(2002); H. Zhang,
        \Journal{\JPG}{30}{S577}{2004}.
\bibitem{antai} A. Tai, \Journal{\JPG}{30}{S809-S818}{2004}; nucl-ex/0404029.
\bibitem{phenix200e}PHENIX Collaboration, S.S. Adler, \etal, nucl-ex/0409028. 
\bibitem{alexi}A.A.P. Suaide, \Journal{\JPG}{30}{S1179-S1182}{2004}; nucl-ex/0404019.
\bibitem{startof1} J. Adams {\it et al.}, (STAR Collaboration), nucl-ex/0309012.
\bibitem{johnson} J. Adams {\it et al.}, (STAR Collaboration), nucl-ex/0401008.
\bibitem{ralfa}R. Averbeck, \Journal{\JPG}{30}{S943-S950}{2004}.
\bibitem{teaney}D. Teaney, BNL seminar; private communications. 
\bibitem{pythia1} T. Sj\"ostrand, L. L\"onnblad and S. Mrenna,
hep-ph/0108264. In this paper we used PYTHIA 6.152 with CTEQ5M1. 
\bibitem{cosmic}I.V. Rakobolskaya {\it et al.},
\Journal{\NPB}{112}{353c}{2003}.
\bibitem{starnch}J. Adams \etal, (STAR Collaboration),
\Journal{\PRL}{91}{172302}{2003};\Journal{\PRL}{91}{072304}{2003}.
\end{thebibliography}
\end{document}